# A robust method for calculating plasma waves absorption in magnetized plasmas and its implementation in the BORAY ray-tracing code

Wanying Yu[a], Huasheng Xie[b,c,*], Aohua Mao[d,e,f*], Haojie Ma[g], Zhengxiong Wang[a,*]

[a] Key Laboratory of Materials Modification by Laser, Ion, and Electron Beams of the Ministry of Education, School of Physics, Dalian University of Technology, Dalian 116024, China

[b] Hebei Key Laboratory of Compact Fusion, Langfang065001, China

[c] ENN Science and Technology Development Co., Ltd., Langfang 065991, China

[d] School of Physics, Harbin Institute of Technology, Harbin 150001, China

[e] Laboratory for Space Environment and Physical Sciences, Harbin 150001, China

[f] China-Russia Belt and Road Joint Laboratory on Advanced Energy and Power Technology, Harbin Institute of Technology, Harbin 150001, China

[g] School of Liberal Arts and Sciences, North China Institute of Aerospace Engineering, Langfang 065000, People's Republic of China

* Corresponding author, E-mail address: huashengxie@gmail.com; aohuamao@hit.edu.cn; zxwang@dlut.edu.cn;

## Abstract

This paper presents a robust numerical method for calculating the total absorption rate of electromagnetic waves in magnetized plasmas, capable of determining the absorption ratio among different plasma components. The method adopts Rönnmark's expressions for the plasma dispersion function, $Z(\zeta)$, and the dielectric tensor, $\boldsymbol{K}$, to overcome the convergence issues and computational inefficiency of the traditional Bessel function summation approach for evaluating the hot plasma dispersion relation, $\boldsymbol{D}$, particularly at large $k_\perp$. It has been implemented and validated across multiple frequency regimes including ion cyclotron (ICRF), lower hybrid (LHRF), and electron cyclotron (ECRF) ranges of frequencies with results benchmarked against conventional $Z(\zeta)$ and Bessel function expansions. Integrated into the BORAY ray-tracing code, the method uses expressions derived from the anti-Hermitian part of the dielectric tensor to compute absorption ratios. This work extends the ray-tracing framework, providing a more reliable tool for wave heating simulations.

# 1. Introduction

Due to the presence of collisionless damping and collisional effects, the plasma waves have been recognized as important approaches for heating plasma and driving current in magnetized confinement fusion research[1-3]. In the research of RF wave heating, the methods employed include the full-wave method and the geometrical optics approximation which yields the ray tracing equations, with the full-wave method being the primary one. The main codes are TASK[4], PICES[5], TORIC[6], etc. However, the full-wave method requires a large amount of computation and the process is extremely complex. The geometric optics method is straightforward and is a commonly used research method for studying the propagation of waves in plasma. Particularly for ECW[7]and LHW[8], excellent results have been obtained using the geometric optics method. For fast waves, due to their relatively long wavelengths, this method is not very suitable for studying the propagation of fast waves in current small - scale experimental devices. Although some researchers have used this method to study the propagation of fast waves in plasma, the results are not satisfactory. Nevertheless, with the progress of magnetic- confinement controlled nuclear fusion research, as the size of tokamak fusion devices increases, and the confinement magnetic field and plasma temperature rise, the geometric optics method can be used to study the propagation of fast waves in such large - scale fusion devices, and certain achievements have been made[9]. In conclusion, it is entirely feasible to use the geometric optics method to study the propagation and absorption of waves.

The dispersion relation for plasma waves describes the connection between the frequency ($\omega$) of a collective perturbation and its wave vector ($k$). This relation is fundamental for studying wave propagation, amplification, attenuation, and plasma instabilities. Its solution requires the computation of the plasma dispersion function $Z(\zeta)$ and the summation of an infinite series of Bessel functions $\Gamma_n$. A key challenge, however, lies in the computational difficulties posed by the traditional integral form of $Z(\zeta)$, as well as poor convergence and slow computation that result from truncating the Bessel function series. Weiss [10] and Qin [11] derived methods to avoid truncated summation of Bessel functions, but these require the use of complex-order Bessel functions, which are computationally expensive. Notably, in space physics applications, Rönnmark provided a method for calculating the kinetic dispersion tensor with superior convergence and computational speed[12]. Therefore, this study adopts Rönnmark's non-relativistic kinetic dispersion relation (KDR), expressed as $D(\omega, \boldsymbol{k}) = \left| \boldsymbol{K}(\omega, \boldsymbol{k}) + (\boldsymbol{k}\boldsymbol{k} - k^2 \boldsymbol{I}) \frac{c^2}{\omega^2} \right| = 0$, where by directly solving for the perpendicular complex wave number $k_\perp = k_{\perp r} + i k_{\perp i}$, the imaginary part $k_{\perp i}$ can be obtained. The absorption rate is then determined by integrating along the ray path[12-14]. Our results demonstrate that this method is robust for Maxwellian distribution plasmas.

This method has been integrated into the BORAY ray-tracing code, with adds the function of calculating the absorption contribution from individual particle species. This allows for more precise analysis of the physical mechanisms of wave energy dissipation. Currently, mainstream ray-tracing codes like CURRAY, TORAY, and GENRAY have proven effective in specific contexts. CURRAY is specifically designed for the lower hybrid/ion cyclotron (LH/IC) frequency range[9], while TORAY's expertise lies in the electron cyclotron (EC) range, leveraging a relativistic FLR-corrected dispersion relation[15]. GENRAY stands out for its broad support of various wave types[16, 17]. While these tools excel in their intended applications, our work aims for greater universality. Our present code implements a unified dispersion relation and a consistent absorption model, allowing it to handle computations across all frequency ranges with a single, streamlined setup. This design offers distinct advantages in generality and user-friendliness.

This paper is organized into four sections. Section 2 is devoted to the theory and methodology, which encompass the kinetic dispersion relation, Rönnmark's formulation, and the method for computing the absorption ratio. Section 3 presents and benchmarks the test cases against results from GENRAY. Finally, a summary and discussion of the work are provided in Section 4.

## 2. Theory and Methods

The formulation proposed by Rönnmark[12] is adopted to mitigate the convergence issues associated with solving the traditional kinetic dispersion relation. Elaborated upon in this section are the general form of the kinetic dispersion relation, the introduction and merits of Rönnmark's formulation, and the expressions for the absorption ratio calculation of each particle species.

## 2.1. Kinetic dispersion relation

We consider an infinite and uniform system in which each particle species, denoted by the index s, possesses a Maxwellian velocity distribution function given by $f_{s0} = \pi^{-3/2} v_{ts}^{-3} \exp\left[-\frac{v_\parallel^2 + v_\perp^2}{v_{ts}^2}\right]$. The background magnetic field is assumed to be uniform, $\boldsymbol{B_0} = (0,0,B_0)$, and the wave vector is $\boldsymbol{k} = (k_x, 0, k_z) = (k\sin\theta, 0, k\cos\theta)$, such that $k_\parallel = k_z$ and $k_\perp = k_x$. The system comprises a total of S particle species, indexed by $s = 1, 2, \ldots, S$. For each species, the charge, mass, density, and temperature are denoted by $q_s$, $m_s$, $n_{s0}$, and $T_{s0}$, respectively. The thermal velocity is defined as $v_{ts} = \sqrt{\frac{2k_B T_{s0}}{m_s}}$, where $k_B$ is the Boltzmann constant.

The KDR is given by:

$$D(\omega, \boldsymbol{k}) = \left|K(\omega, \boldsymbol{k}) + (\boldsymbol{kk} - k^2 \boldsymbol{I})\frac{c^2}{\omega^2}\right| = 0, \tag{1}$$

where,

$$\boldsymbol{K} = \boldsymbol{I} + \boldsymbol{Q} = \boldsymbol{I} - \frac{\sigma}{i\omega\epsilon_0}, \quad \boldsymbol{Q} = -\frac{\sigma}{i\omega\epsilon_0}. \tag{2}$$

Define $a_s = k_\perp \rho_{cs}$, $b_s = a_s^2 = k_\perp^2 \rho_{cs}^2$, $\rho_{cs} = v_{ts}/\sqrt{2}\omega_{cs}$, $\boldsymbol{n} = \boldsymbol{k}c/\omega$, $\omega_{cs} = q_s B_0/m_s$, $\omega_{ps} = \sqrt{n_s q_s^2/\epsilon_0 m_s}$, $c = 1/\sqrt{\mu_0 \epsilon_0}$, $\zeta_{sn} = (\omega - n\omega_{cs})/k_z v_{ts}$, where $c$ is the speed of light, $\omega_{ps}$ and $\omega_{cs}$ are the plasma frequency and cyclotron frequency for each species, $\epsilon_0$ is the permittivity of free space, and $\mu_0$ is the permeability of free space. Following standard derivations[18-20], we obtain:

$$\boldsymbol{K} = \boldsymbol{I} + \sum_s \frac{\omega_{ps}^2}{\omega^2}[\sum_{n=-\infty}^{\infty} \zeta_{s0} Z(\zeta_{sn}) \boldsymbol{X}_{sn} + 2\zeta_{s0}^2 \hat{\boldsymbol{z}}\hat{\boldsymbol{z}}], \tag{3}$$

where,

$$\boldsymbol{X}_{sn} = \begin{bmatrix} \frac{n^2}{b_s}\Gamma_{sn} & in\Gamma'_{sn} & \sqrt{2}\zeta_{sn}\frac{n}{a_s}\Gamma_{sn} \\ -in\Gamma'_{sn} & \frac{n^2\Gamma_{sn}}{b_s} - 2b_s\Gamma'_{sn} & -i\sqrt{2}\zeta_{sn}a_s\Gamma'_{sn} \\ \sqrt{2}\zeta_{sn}\frac{n}{a_s}\Gamma_{sn} & i\sqrt{2}\zeta_{sn}a_s\Gamma'_{sn} & 2\zeta_{sn}^2\Gamma_{sn} \end{bmatrix}. \quad (4)$$

Here, $Z(\zeta) = \frac{1}{\sqrt{\pi}}\int_{-\infty}^{+\infty}\frac{e^{-z^2}}{z-\zeta}dz$ is the plasma dispersion function [21, 22], $\Gamma_{sn} \equiv \Gamma_n(b_s)$, $\Gamma'_{sn} \equiv \Gamma'_n(b_s)$, with $\Gamma_n(b) = I_n(b)e^{-b}$, $\Gamma'_n(b) = (I'_n - I_n)e^{-b}$, $I'_n(b) = (I_{n+1} + I_{n-1})/2$, $I_{-n} = I_n$, $Z'(\zeta) = -2[1 + \zeta Z(\zeta)]$, and $I_n$ is the modified Bessel function of the first kind of order n. Note that $K_{yx} = -K_{xy}$, $K_{zx} = K_{xz}$, $K_{zy} = -K_{yz}$.

The computation of the dielectric tensor above involves the complex integral form of the plasma dispersion function and an infinite series summation over Bessel functions (typically truncated, requiring determination of the truncation value for n). This leads to numerical difficulties, significant computational cost, and convergence issues. Therefore, it is necessary to employ an efficient rational function approximation method to simplify the expression of the dielectric tensor, ensuring accuracy while significantly improving computational efficiency.

Rönnmark's method well satisfies these requirements. This method uses Padé approximation and J-pole expansion for the dispersion function $Z(\zeta)$, expressed as $Z(\zeta) \simeq \sum_{j=1}^{J}\frac{b_j}{\zeta-c_j}$, achieving errors less than $10^{-6}$ even for J=8 [12]. Higher-order expansion coefficients for improved accuracy are also provided in [23]. In this method, the tensor Q in the KDR is modified to:

$$Q = \sum_s \frac{\omega_{ps}^2}{\omega\omega_{cs}}\sum_j b_j \cdot \begin{bmatrix} R_{sj} & \frac{i}{x_{sj}}R'_{sj} & \frac{\sqrt{2}a_sc_j}{x_{sj}}R_{sj} \\ -\frac{i}{x_{sj}}R'_{sj} & R_{sj} - \frac{2b_s}{x_{sj}^2}R'_{sj} & -\frac{i\sqrt{2}a_sc_j}{x_{sj}^2}R'_{sj} \\ \frac{\sqrt{2}a_sc_j}{x_{sj}}R_{sj} & \frac{i\sqrt{2}a_sc_j}{x_{sj}^2}R'_{sj} & \frac{2c_j^2}{x_{sj}^2}(x_{sj} + b_sR_{sj}) \end{bmatrix}, \quad (5)$$

where $b_j$ and $c_j$ are the coefficients for the Padé approximation of the $Z(\zeta)$ function [12, 21], $x_{sj} = (\omega - k_zv_{ts}c_j)/\omega_{cs}$, and $R_{sj} = R(x_{sj}, b_s)$. Here,

$$R(x,\lambda) \equiv \sum_{n=-\infty}^{\infty}\frac{n^2\Gamma_n(\lambda)}{\lambda(x-n)} = -\frac{x}{\lambda} + x^2\sum_{n=-\infty}^{\infty}\frac{\Gamma_n(\lambda)}{\lambda(x-n)}, \quad (6)$$

$$R'(x,\lambda) \equiv \frac{\partial[\lambda R(x,\lambda)]}{\partial\lambda} = \sum_{n=-\infty}^{\infty}\frac{n^2\Gamma'_n(\lambda)}{(x-n)} = x^2\sum_{n=-\infty}^{\infty}\frac{\Gamma'_n(\lambda)}{(x-n)}, \quad (7)$$

with the properties $\sum_{n=-\infty}^{\infty} \Gamma_n(b) = 1$, $\sum_{n=-\infty}^{\infty} n\Gamma_n(b) = 0$, $\sum_{n=-\infty}^{\infty} n^2 \Gamma_n(b) = b$. The computations of R and R' employ a fast and convergent scheme provided by Rönnmark in his WHAMP code[12]. We compared this method with the traditional one and found it to be computationally faster, with the advantage becoming more pronounced when larger N values are required to ensure convergence.

Based on the above dispersion relation, we can solve for $\omega$ and $k$. The total absorption rate along the ray path is then obtained by integrating the imaginary part, using the following equation[18],

$$P(s) = P_0 \cdot e^{-2\int_0^s k_i \cdot dr} = P_0 \cdot e^{-2\int_0^t \omega_i \cdot dt}. \tag{8}$$

## 2.2. Absorption ratio by species

Starting from Maxwell's equations, we derive Poynting's theorem:

$$\frac{c}{4\pi} \nabla \cdot (\mathbf{E} \times \mathbf{B}) = -\frac{1}{4\pi}\left(\mathbf{B} \cdot \frac{\partial \mathbf{B}}{\partial t} + \mathbf{E} \cdot \frac{\partial \mathbf{D}}{\partial t}\right), \tag{9}$$

which is interpreted as a conservation theorem. The Poynting vector $c \cdot \mathbf{E} \times \mathbf{B}/4\pi$ represents the flux of electromagnetic energy, while the term on the right-hand side represents the rate of change of energy density. Using time-averaged relations [24]:

$$\nabla \cdot \mathbf{P} + \frac{\partial W}{\partial t} = 0, \tag{10}$$

$$\mathbf{P} \equiv \frac{c}{16\pi}(\mathbf{E}_1^* \times \mathbf{B}_1 + \mathbf{E}_1 \times \mathbf{B}_1^*) e^{2\phi_i(r,t)}, \tag{11}$$

$$\frac{\partial W}{\partial t} \equiv \frac{1}{16\pi}[2\omega_i \mathbf{B}_1^* \cdot \mathbf{B}_1 + \omega_i^* \mathbf{E}_1^* \cdot (\boldsymbol{\epsilon} + \boldsymbol{\epsilon}^\dagger) \cdot \mathbf{E}_1 + \omega_r \mathbf{E}_1^* \cdot (-i\boldsymbol{\epsilon} + i\boldsymbol{\epsilon}^\dagger) \cdot \mathbf{E}_1]e^{2\phi_i(r,t)}, \tag{12}$$

where $\boldsymbol{\epsilon}^\dagger$ is the Hermitian conjugate of $\boldsymbol{\epsilon}$:

$$\boldsymbol{\epsilon}^\dagger(\omega, \mathbf{k}) = [\tilde{\boldsymbol{\epsilon}}(\omega, \mathbf{k})]^*, \tag{13}$$

i.e., the conjugate of the transpose matrix (the transpose tensor $\widetilde{\epsilon_{ij}} = \epsilon_{ji}$ has the property $\mathbf{A} \cdot \tilde{\boldsymbol{\epsilon}} \cdot \mathbf{B} = \mathbf{B} \cdot \boldsymbol{\epsilon} \cdot \mathbf{A}$ for all $\mathbf{A}$ and $\mathbf{B}$. In this notation, the Hermitian conjugate is $\epsilon_{ij}^\dagger = \epsilon_{ji}^*$, or $\mathbf{A} \cdot \boldsymbol{\epsilon}^\dagger \cdot \mathbf{B} = \mathbf{B} \cdot \boldsymbol{\epsilon}^* \cdot \mathbf{A}$).

The sum of the terms on the right-hand side of Eq. (13) accounts for both overcoming dissipative losses and forming the rate of wave energy consumption. The growth/decay of the wave amplitude can be represented by a complex frequency $\omega = \omega_r + \omega_i$, where $\omega_i \neq 0$. On the other hand, under steady-state conditions when $\partial W/\partial t = 0$ (i.e., $\omega_i = 0$), the plasma is lossless. Equation (12) directly indicates that if $\epsilon(\omega_r)$ is Hermitian ($\epsilon = \epsilon^\dagger$), then the plasma is lossless. Furthermore, if the plasma is lossless for all $\boldsymbol{E}_1$, then $\epsilon(\omega_r)$ must be Hermitian.

Generally, $\epsilon(\omega_r)$ is not entirely lossless. We can separate its lossless part using the above formulation. The separation is performed at $\omega_r, k_r$, where the interpretation of $\partial W/\partial t$ in Eq. (12) is particularly clear. Denote the Hermitian and anti-Hermitian parts of $\epsilon(\omega_r, \boldsymbol{k}_r)$ as $\epsilon_h$ and $i\epsilon_a$, respectively:

$$\epsilon_h(\omega_r, \boldsymbol{k}_r) = \frac{1}{2}[\epsilon(\omega_r, \boldsymbol{k}_r) + \epsilon^\dagger(\omega_r, \boldsymbol{k}_r)], \tag{14}$$

$$\epsilon_a(\omega_r, \boldsymbol{k}_r) = \frac{1}{2}[\epsilon(\omega_r, \boldsymbol{k}_r) - \epsilon^\dagger(\omega_r, \boldsymbol{k}_r)]. \tag{15}$$

For waves with varying amplitude ($\omega_i \neq 0$ or $k_i \neq 0$), $\epsilon_h$ itself can generate an anti-Hermitian component. By analytically continuing $\epsilon_h(\omega_r, \boldsymbol{k}_r)$ a small distance in the $\omega, \boldsymbol{k}$ plane:

$$\epsilon_h(\omega_r + i\omega_i, \boldsymbol{k}_r + i\boldsymbol{k}_i) = \left[\epsilon_h + i\omega_i \frac{\partial}{\partial \omega}\epsilon_h + i\boldsymbol{k}_i \cdot \frac{\partial}{\partial \boldsymbol{k}}\epsilon_h\right]_{\omega_r, \boldsymbol{k}_r} + \cdots,$$

$$\epsilon_h^\dagger(\omega_r + i\omega_i, \boldsymbol{k}_r + i\boldsymbol{k}_i) = \left[\epsilon_h - i\omega_i \frac{\partial}{\partial \omega}\epsilon_h - i\boldsymbol{k}_i \cdot \frac{\partial}{\partial \boldsymbol{k}}\epsilon_h\right]_{\omega_r, \boldsymbol{k}_r} + \cdots, \tag{16}$$

$$-i\epsilon_h(\omega, \boldsymbol{k}) + i\epsilon_h^\dagger(\omega, \boldsymbol{k}) = 2\omega_i \frac{\partial}{\partial \omega}\epsilon_h(\omega_r, \boldsymbol{k}_r) + 2\boldsymbol{k}_i \cdot \frac{\partial}{\partial \boldsymbol{k}}\epsilon_h(\omega_r, \boldsymbol{k}_r) + \cdots, \tag{17}$$

The vector dot products in Eqs. (16) and (17) are between $\boldsymbol{k}_i$ and $\partial/\partial \boldsymbol{k}$, where $\partial/\partial \boldsymbol{k} = \hat{\boldsymbol{x}}\partial/\partial k_x + \hat{\boldsymbol{y}}\partial/\partial k_y + \hat{\boldsymbol{z}}\partial/\partial k_z$. Now, for $|\omega_i| \ll |\omega|$, $|\boldsymbol{k}_i| \ll |\boldsymbol{k}|$, we can recast $\partial W/\partial t$ in Eq. (12) using Eqs. (15) and (17):

$$\frac{\partial W}{\partial t} = \frac{1}{8\pi}[\omega_i \boldsymbol{B}_1^* \cdot \boldsymbol{B}_1 + \omega_i \boldsymbol{E}_1^* \cdot \epsilon_h \cdot \boldsymbol{E}_1 + \omega_r \boldsymbol{E}_1^* \cdot \left(\epsilon_a + \omega_i \frac{\partial}{\partial \omega}\epsilon_h + \boldsymbol{k}_i \cdot \frac{\partial}{\partial \boldsymbol{k}}\epsilon_h\right) \cdot \boldsymbol{E}_1]e^{2\phi_i} \tag{18}$$

where, $\epsilon_h = \epsilon_h(\omega_r, \boldsymbol{k}_r)$ and $\epsilon_a = \epsilon_a(\omega_r, \boldsymbol{k}_r)$. We see that the amplitude variation (terms involving $\omega_i, k_i$ in Eq. (18)) affects the wave energy density in several ways: firstly, it changes the quantity $(B^2 + \boldsymbol{E} \cdot \boldsymbol{D})/8\pi$. But since the amplitude variation involves a spectrum of frequencies and possibly wavelengths, contributions from the derivatives of $\epsilon(\omega, \boldsymbol{k})$, namely $\partial \epsilon(\omega, \boldsymbol{k})/\partial \omega$ and $\partial \epsilon(\omega, \boldsymbol{k})/\partial \boldsymbol{k}$, also appear.

Finally, the divergence of the Poynting vector according to Eq. (11) is:

$$\nabla \cdot \boldsymbol{P} = \frac{c}{16\pi} \nabla \cdot [(\boldsymbol{E}_1^* \times \boldsymbol{B}_1 + \boldsymbol{E}_1 \times \boldsymbol{B}_1^*) \, exp(2\phi_i(\boldsymbol{r},t))] = -2\boldsymbol{k}_i \cdot \boldsymbol{P}. \tag{19}$$

Poynting's theorem can be expressed as:

$$\nabla \cdot \boldsymbol{P} + \frac{\partial W}{\partial t} = -2\boldsymbol{k}_i \cdot (\boldsymbol{P} + \boldsymbol{T}) + 2\omega_i W + \left.\frac{\partial W}{\partial t}\right|_{lossy} = 0, \tag{20}$$

where,

$$\boldsymbol{P} = \frac{c}{16\pi}(\boldsymbol{E}^* \times \boldsymbol{B} + \boldsymbol{E} \times \boldsymbol{B}^*), \tag{21}$$

$$\boldsymbol{T} = -\frac{\omega}{16\pi} \boldsymbol{E}^* \cdot \frac{\partial}{\partial \boldsymbol{k}} \epsilon_h \cdot \boldsymbol{E}, \tag{22}$$

$$W = \frac{1}{16\pi}\left[\boldsymbol{B}^* \cdot \boldsymbol{B} + \boldsymbol{E}^* \cdot \frac{\partial}{\partial \omega}(\omega \epsilon_h) \cdot \boldsymbol{E}\right], \tag{23}$$

$$\left.\frac{\partial W}{\partial t}\right|_{lossy} = \frac{\omega_r}{8\pi} \boldsymbol{E}^* \cdot \epsilon_a \cdot \boldsymbol{E}. \tag{24}$$

From the above formulas, the absorption ratio for species s relative to the total absorption is given by:

$$\boldsymbol{P}_{tot} = \frac{\left.\frac{\partial W}{\partial t}\right|_{lossy(s)}}{\left.\frac{\partial W}{\partial t}\right|_{lossy(sum)}} = \frac{\boldsymbol{E}^* \cdot \epsilon_a^s \cdot \boldsymbol{E}}{\boldsymbol{E}^* \cdot \epsilon_a^{sum} \cdot \boldsymbol{E}}, \tag{25}$$

where $\epsilon_a^s$ is the anti-Hermitian part of the dielectric tensor contribution from species s, and $\epsilon_a^{sum} = \sum_s \epsilon_a^s$.[25] The calculation of $\epsilon_a^s$ follows the method described in the previous subsection. It should be noted that both the input $\omega$ and **k** are real numbers.

## 3. Numerical implementation and verification

This section first compares the two dispersion relation solution algorithms. Then, using GENRAY as a benchmark, it verifies the consistency of total power absorption across a broad frequency range. Finally, it analyzes the distribution of absorption contributions among particle species, aiming to comprehensively validate the reliability and robustness of the present model.

### 3.1. Comparison of dispersion relation solvers

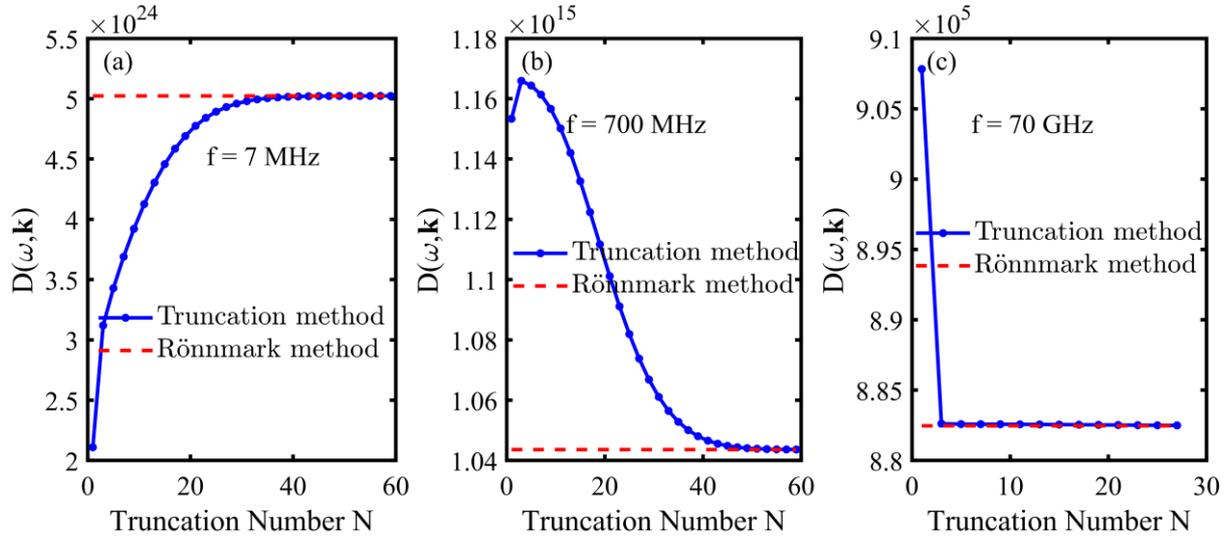

**Figure 1.** Comparison between results obtained using the Bessel function truncation summation method and Rönnmark's method for large $k_\perp$. The comparison is shown for three different frequency range: (a) ICRF (7 MHz), (b) LHRF, (c) ECRF.

Figure 1 shows a comparison of results obtained using the Bessel function truncation method and Rönnmark's method for large $k_\perp$ in three different frequency range under identical parameters, illustrating the variation of $D(\omega, \mathbf{k})$ with the truncation number N. It can be observed that for a 7 MHz ICW (a), the Bessel function truncation method only gradually converges for larger N. Before convergence is reached, the error of this method can be as high as 60%. For the LHW (b), convergence also requires a large N value. For the ECW (c), convergence is achieved with a relatively small N. However, determining a suitable N value is difficult, and choosing an excessively large N reduces computational efficiency. In contrast, Rönnmark's method guarantees convergent results without specifying an N value, offering significant advantages in computational efficiency and stability.

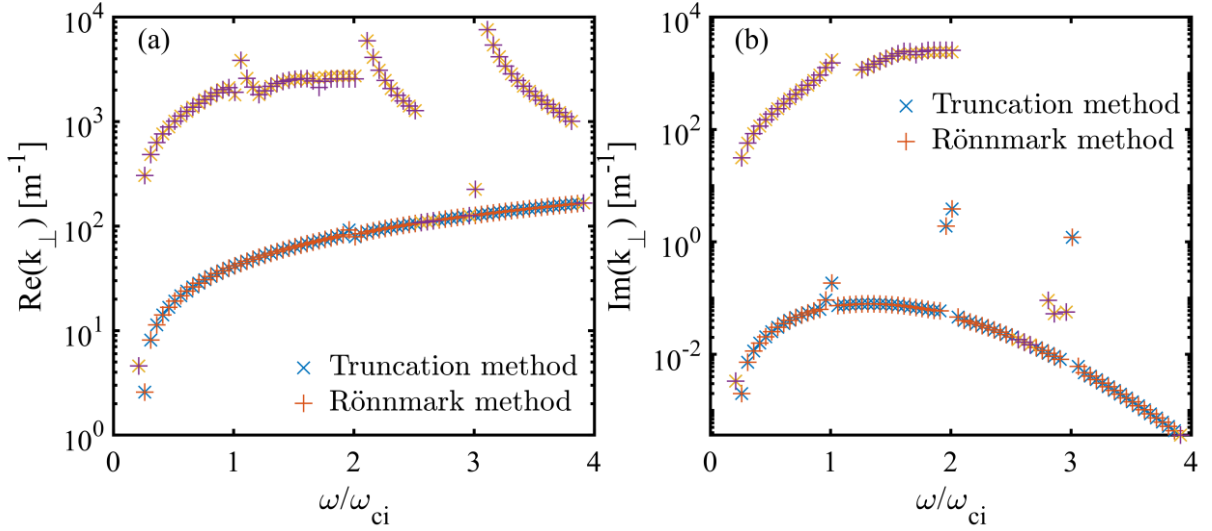

**Figure 2.** Comparison between truncation method (Bessel function truncation summation) and Rönnmark's method. Results show excellent agreement for both (a) Re($k_\perp$) and (b) Im($k_\perp$) of the predominant modes.

The complex vertical wavenumber $k_\perp$, a critical parameter for subsequent wave absorption calculations, is directly computed using two distinct methods: a truncation method based on Bessel function summation and Römmark's method. As depicted in Figure 2, the results for both (a) the real part and (b) the imaginary part of $k_\perp$ exhibit excellent agreement. This strong consistency not only validates the accuracy of both approaches but also provides a reliable foundation for the following analysis of wave absorption.

### 3.2. Wave absorption

The reliability and robustness of the present model are substantiated through a three-fold validation in this section. The analysis first compares the two solution algorithms for the dispersion relation, then benchmarks the total power absorption against GENRAY across a wide frequency range, and finally delineates the absorption contribution from each particle species.

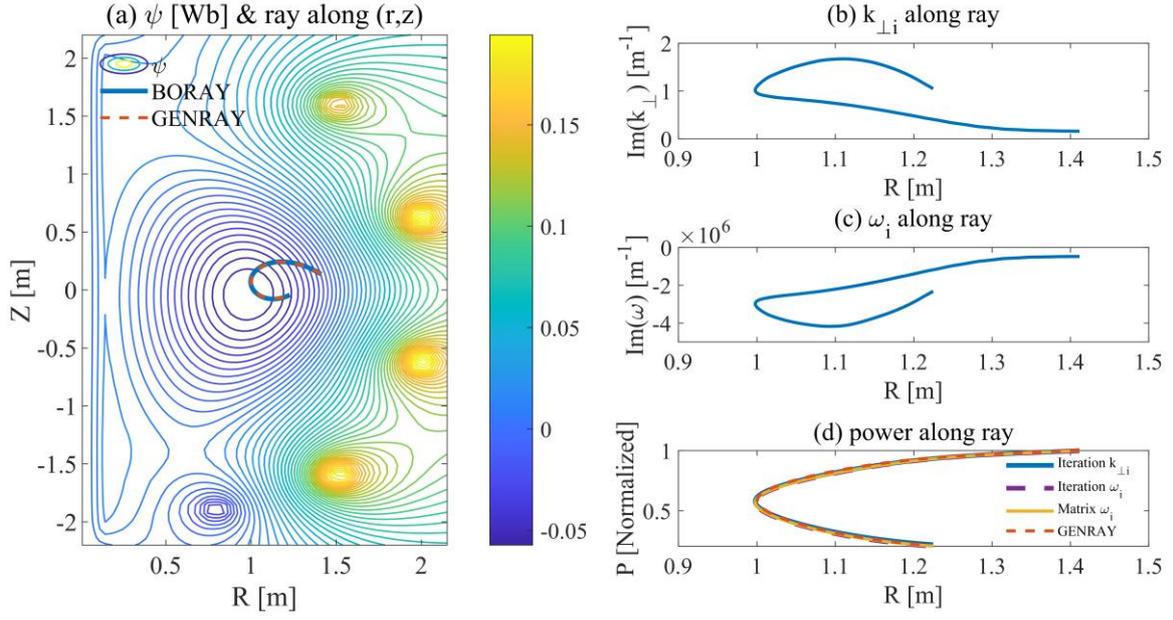

**Figure 3.** (a) Ray trace and (d) total absorption comparison for a 30 MHz ICRF wave in the NSTX device. Here, three methods are adopted: solving for $\text{Im}(k_\perp)$; solving for $\text{Im}(\omega)$; and using the matrix method to determine $\text{Im}(\omega)$. All methods show good agreement with GENRAY.

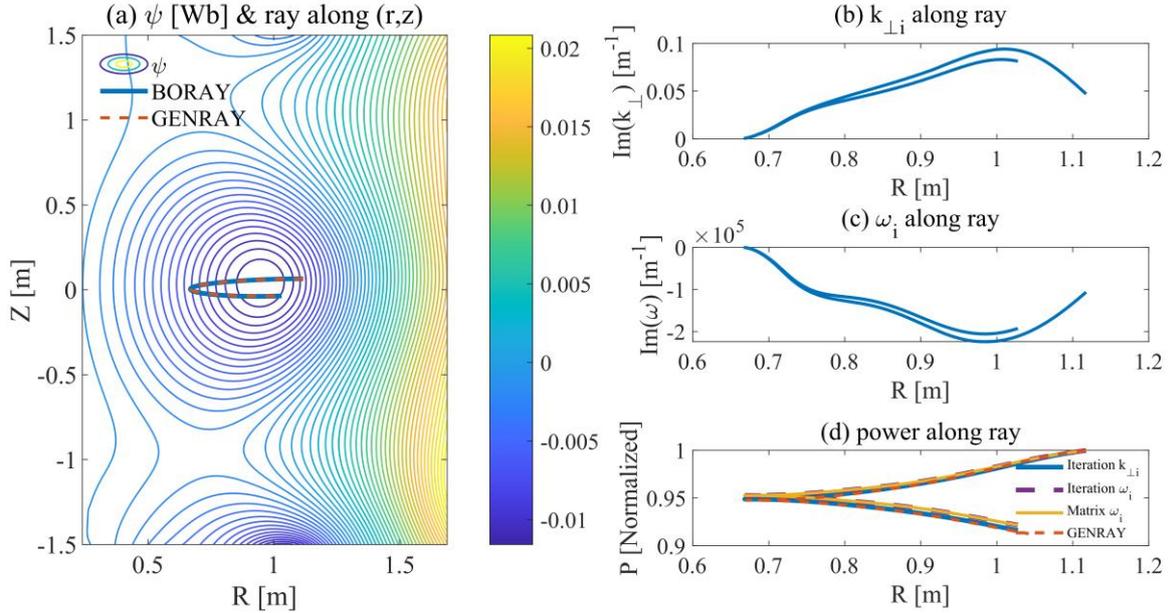

**Figure 4.** (a) Ray trace and (d) total absorption comparison for an 8 MHz ICRF wave in the EXL-50U device. Method 1 shows the closest agreement with GENRAY; the other methods also agree reasonably well.

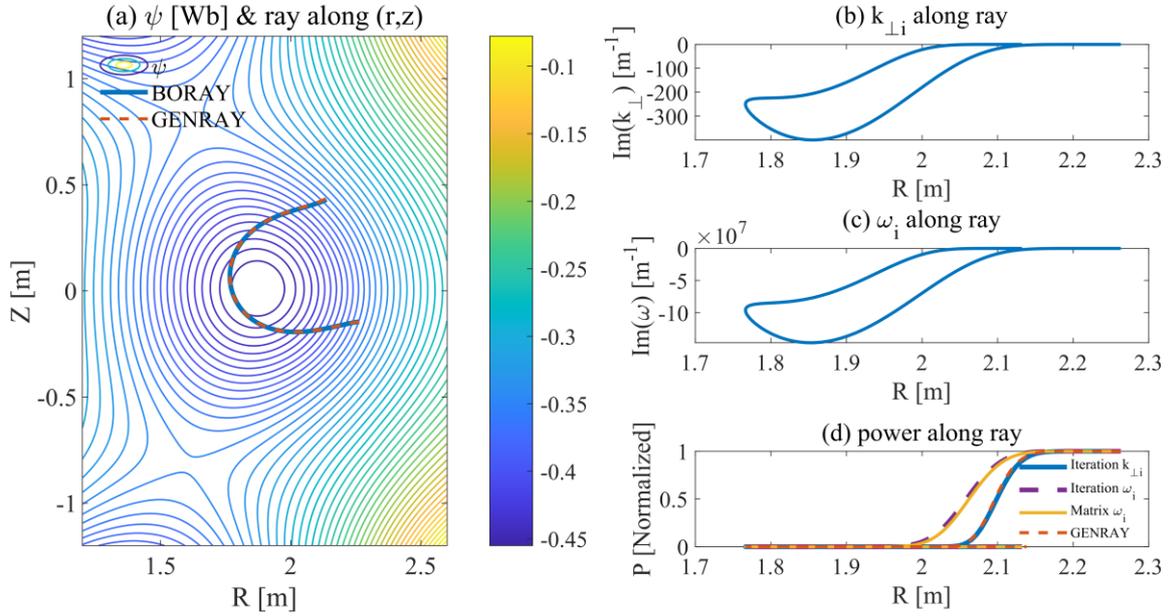

**Figure 5.** (a) Ray trace and (d) total absorption comparison for a 100 GHz electron cyclotron wave (X-mode) in the EAST device. All methods show good agreement with GENRAY.

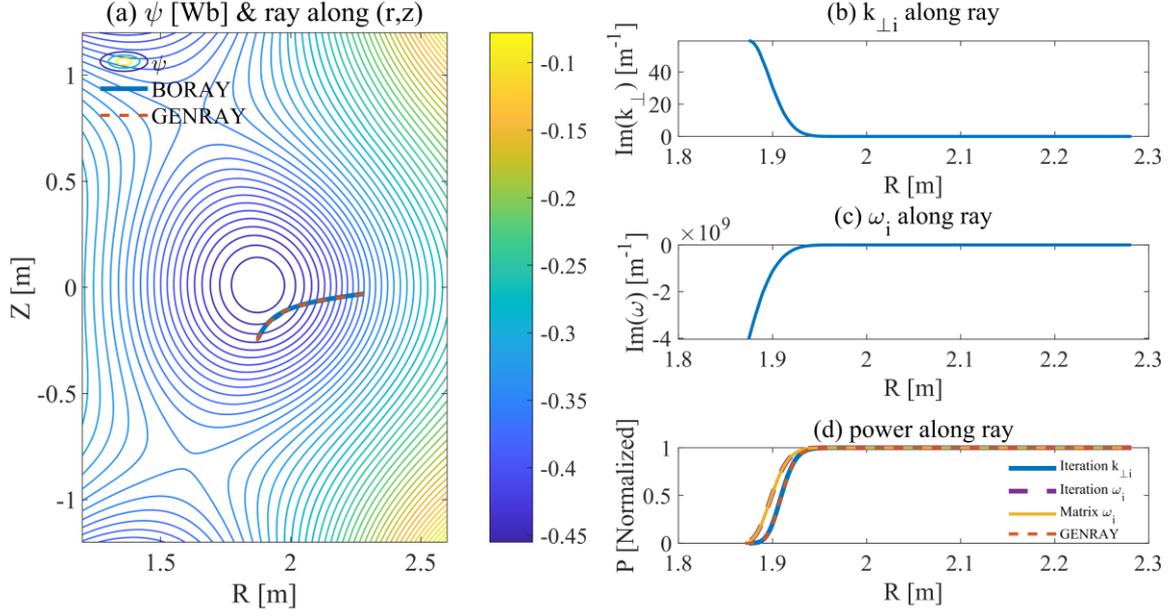

**Figure 6.** (a) Ray trace and (d) total absorption comparison for a 2.5 GHz lower hybrid wave in the EAST device. Method 1 shows the closest agreement with GENRAY; the absorption calculated by the other methods shows larger discrepancies compared to GENRAY.

In this subsection, we demonstrated the performance comparison between several absorption calculation methods in BORAY and GENRAY under standard tokamak and steady-state operating conditions, covering high to low frequency ranges: electron cyclotron waves (ECW, X-mode), lower hybrid waves (LHW), and ion cyclotron waves (ICW).

Figure 3 shows the comparison for a 30 MHz ICW in the NSTX tokamak ($B_0 = 0.3$ T, major radius $R_0 = 0.85$ m, safety factor $q_0 = 1.3786$, density $n_{e0} = 2 \times 10^{19}$ m$^{-3}$, temperature $T_{e0} = T_{i0} = 2500$ eV).

Figure 4 shows the comparison for an 8 MHz ICW in the EXL-50U tokamak ($B_0 = 0.7560$ T, $R_0 = 0.9524$ m, $q_0 = 0.6838$, $n_{e0} = 2 \times 10^{19}$ m$^{-3}$, $T_{e0} = T_{i0} = 10000$ eV).

Figure 5 shows the comparison for a 100 GHz ECW (X-mode) in the EAST tokamak ($B_0 = 1.78$ T, $R_0 = 1.88$ m, $q_0 = 1.5$, $n_{e0} = 5 \times 10^{19}$ m$^{-3}$, $T_{e0} = T_{i0} = 500$ eV).

Figure 6 shows the comparison for a 2.5 GHz LHW in the EAST tokamak, using the same equilibrium as in Figure 5 but with different density and temperature ($n_{e0} = 1 \times 10^{19}$ $m^{-3}, T_{e0} = T_{i0} = 200$ eV).

In these benchmark cases, the ray trajectories and power absorption agree well with GENRAY. However, when the solutions for ω are close to each other, mode jumping can easily occur. The matrix method is also unstable for the same reason. BORAY provides a derivative-based method to find $\omega_i$, but its applicable conditions are stringent ($D_r \gg D_i$). Considering both stability and implementation difficulty, this study ultimately selected the method of solving for the imaginary part of $k_\perp$ to calculate absorption. Compared to the first two methods, this approach is stable, rarely experiences mode jumping, and currently provides stable power absorption calculations in most situations. In contrast, GENRAY has different built-in absorption models for different waves, requiring manual adjustment of many input parameters when switching frequency range, which is complex and prone to systematic errors due to parameter mismatch. Our method uses the same root-finding framework based on the unified dispersion relation for all frequency range, requiring only minimal input parameters, significantly lowering the barrier to use, and providing a simpler, more stable solution for real-time calculations of multi-band synergistic heating.

### 3.3. Absorption by species

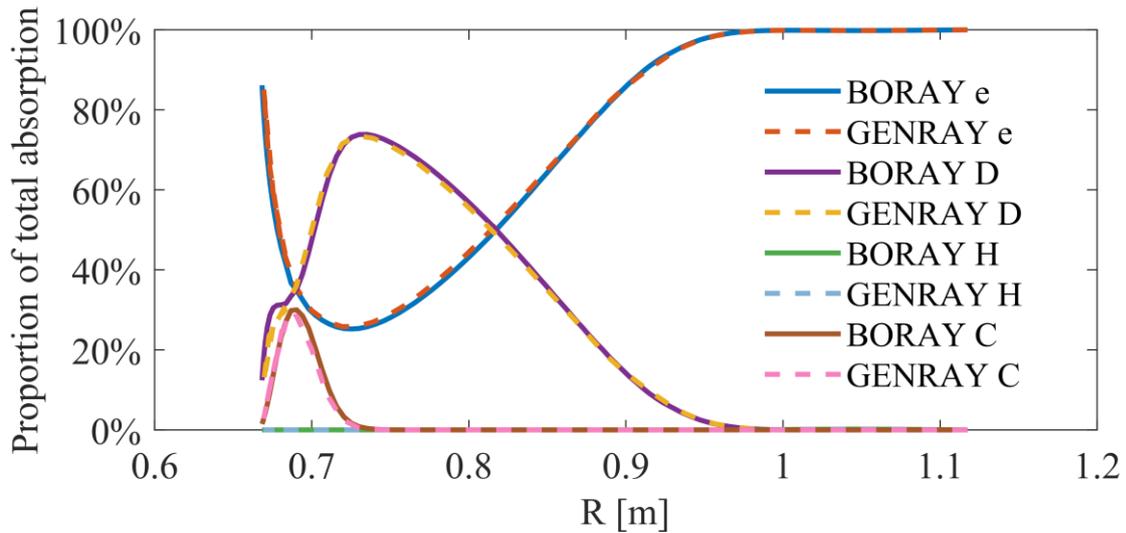

**Figure 7.** Absorption ratio by particle species for an 8 MHz ICW in the EXL-50U device. The equilibrium parameters are consistent with the previous section. Results from BORAY are in basic agreement with GENRAY.

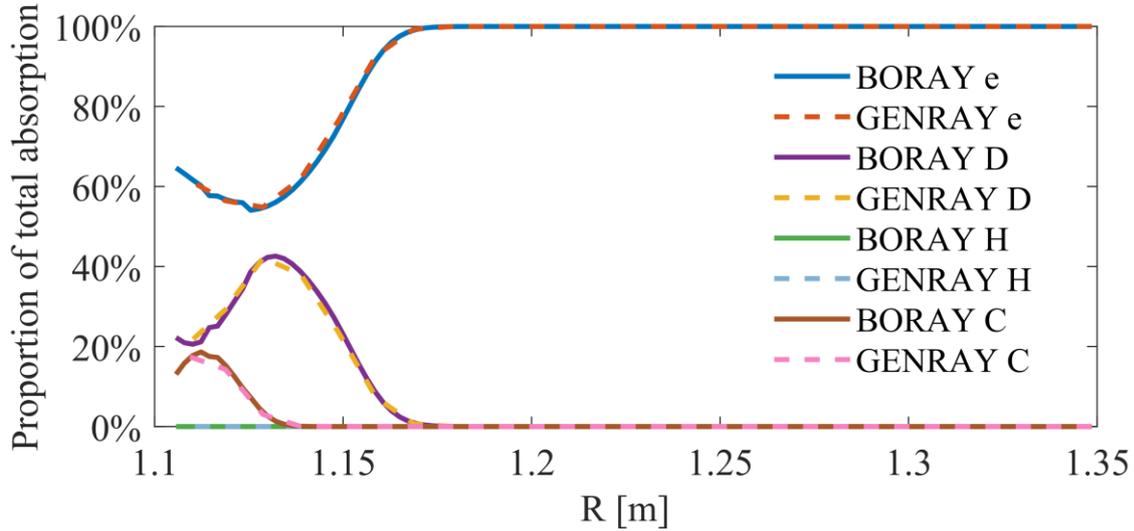

**Figure 8.** Absorption ratio by particle species for a 3 MHz ICW in the NSTX device. The equilibrium parameters are consistent with the previous section; a wave frequency of 3 MHz is used to make the absorption effects of different species more pronounced. Results from BORAY are in basic agreement with GENRAY.

Figures 7 and 8 present the simulation results of ICW absorption by the BORAY code in two distinct tokamak devices. Figure 7 corresponds to an 8 MHz ICW in the EXL-50U device, while Figure 8 depicts a 3 MHz ICW in the NSTX device. This lower frequency was chosen to make the absorption effects of different particle species more pronounced. Both figures break down the absorption ratio by particle species, providing a clear comparison of the relative contributions of various particles to wave energy dissipation. The equilibrium parameters used in these simulations are consistent with the previous section. As shown, the results from BORAY are in essential agreement with those from the established GENRAY code, further validating the effectiveness and reliability of the BORAY model when applied to different devices and parameters.

## 4. Conclusions

A robust numerical method for solving wave absorption problems has been developed and validated in this study, utilizing Rönnmark's expressions. This method not only computes the total wave absorption effectively but also accurately resolves the specific contribution ratios of different particle species (e.g., electrons, different ion species) to energy absorption. The main accomplishments are reflected in the following three aspects:

First, a key implementation is the adoption of Rönnmark's expressions for the dispersion relation, which significantly boosts computational efficiency and stability. This results in a faster, more direct solution process that robustly avoids the numerical pitfalls of traditional methods.

Second, the method provides a distinct advantage in calculating wave absorption. It simplifies the user experience compared to mature codes like GENRAY by generalizing the workflow and removing the need to tune numerous frequency-specific parameters. Extensive verification shows excellent agreement with GENRAY across multiple range (e.g., ICRF, ECRF, LHRF), with all results being numerically stable and free from divergence, conclusively demonstrating the method's reliability.

Finally, the study enables a generalized calculation of the wave energy absorption ratio among different particle species. This achievement offers a powerful analytical tool, allowing researchers to precisely trace dissipation paths and uncover the specific physics of wave-particle interactions.

Notwithstanding these advances, this work is subject to certain limitations. The primary constraint lies in the theoretical model, which is premised on two key assumptions: a homogeneous magnetized plasma and Maxwellian particle velocity distributions. In reality, tokamak plasmas are markedly inhomogeneous and may involve non-equilibrium distribution functions and relativistic effects. These complex physical phenomena fall outside the scope of the present model and represent important directions for future extension.

In summary, this study has established a robust, generalizable, and physically insightful framework for wave absorption calculation. While the current model employs necessary simplifying assumptions, its correctness and practical utility have been substantiated. This work lays a solid foundation for subsequent, more detailed investigations of wave-particle interactions and points to promising avenues for further development.

**Acknowledge**

This work is supported by the National Magnetic Confinement Fusion Energy R&D Program of China (Nos.2022YFE03190400), the State Key Laboratory of Advanced Electromagnetic Technology (Grant No. AET 2025KF024), the Open subject of Key Laboratory of Frontier Physics in Controlled Fusion, the National Natural Science Foundation of China (Grant Nos. 12435014 and 12575228), the Fundament Research Funds for the Central Universities (Grant Nos. DUT24GF110 and DUT25LAB108).